\documentstyle[twocolumn,aps,epsfig]{revtex}
\begin{document}
%
\title{
Correlation functions of higher-dimensional 
Luttinger liquids
}
\author{Lorenz Bartosch and Peter Kopietz}
\address{Department of Physics and Astronomy, University of California,
Los Angeles, 90095\cite{address}\\
Institut f\"{u}r Theoretische Physik der Universit\"{a}t G\"{o}ttingen,
Bunsenstrasse 9, D-37073 G\"{o}ttingen, Germany}
\date{28 April 1998}
\maketitle
\begin{abstract}
Using higher-dimensional bosonization,
we study correlation functions
of fermions
with singular forward scattering. 
Following Bares and Wen [Phys. Rev. B 48, 8636 (1993)],
we consider density-density interactions
in $d$ dimensions  
that diverge in Fourier space for small
momentum transfers $|{\bf q}|$ as $|{\bf q}|^{- \eta}$ with $\eta = 2 (d-1)$.
In this case the single-particle Green's function shows Luttinger
liquid behavior. However, in contrast to $d=1$, in higher dimensions
the singularity of the momentum distribution
$n_{\bf{k}}$ for wave-vectors ${\bf{k}}$
close to the Fermi wave-vector $k_F$ 
is characterized by a {\it{different exponent}}
than the singularity of the density of states  $\rho ( \omega )$
for frequencies $\omega$ close to zero.
We also calculate the irreducible polarization $\Pi_{\star} ( {\bf{q}} , \omega )$
for $| {\bf{q}} | \approx 2 k_F$. 
Whereas in the one-dimensional Tomonaga-Luttinger model
$\Pi_{\star} ( \pm 2 k_F + q , \omega )$ 
exhibits anomalous scaling,  we show that in $d> 1$ the leading singular
corrections cancel.
Thus, even though singular density-density
interactions in $d > 1$ lead to Luttinger
liquid behavior of the single-particle Green's function,
gauge invariant correlation functions
such as the polarization or the conductivity 
show
conventional Fermi liquid behavior.
We discuss consequences for the effect of disorder on higher-dimensional Luttinger liquids.
\end{abstract}
\narrowtext
%
%
\section{Introduction}
\setcounter{section}{1}
\label{sec:intro}
In spite of intense theoretical efforts, 
the question whether correlated electrons in two dimensions
can exhibit 
a non-Fermi liquid (NFL) ground state still has no generally accepted answer.
The solution to this problem may be the key to a proper understanding
of superconductivity in the cuprates. There is little doubt that 
all experiments probing
the normal state 
of the optimally doped cuprates can be consistently interpreted in terms
of NFL behavior\cite{Anderson97}. 
However, from the theoretical side
there exists no agreement on the microscopic mechanism responsible
for the experimentally seen NFL behavior.
The difficulty lies in the fact that a possible breakdown of
the Fermi liquid state
cannot be described by treating the electron-electron interactions
perturbatively.  
Therefore, 
the development of non-perturbative methods in dimensions $d > 1$
is rather important. 

Recently, considerable
progress has been made for the special case that the interaction
between the electrons is dominated by forward scattering.
This means that the typical momentum ${\bf{q}}$ transfered by
the interaction has a magnitude that is
small compared with the Fermi wave-vector
$k_F$. In this case the single-particle Green's function 
$G ( {\bf{k}} , \omega )$
satisfies 
an asymptotic Ward-identity, which can be used to sum 
the most singular terms in the
expansion of
$G ( {\bf{k}} , \omega )$ 
to all orders in the interaction\cite{Metzner97}. 
This Ward identity  forms also the basis\cite{Kopietz97}
for the generalization of the
bosonization method to arbitrary 
dimensions\cite{Kopietz97,Luther79,Kopietz95,Kopietz96,Froehlich95}.

In this work we shall use higher-dimensional bosonization
to study one- and two-particle
correlation functions of higher-dimensional Luttinger liquids.
As a microscopic model system for a NFL in
$d >1$, we shall consider 
spinless fermions interacting with a long-range density-density
interaction $V ( {\bf{r}}  - {\bf{r}}^{\prime} )$ , such that the Fourier transform
$f_{\bf{q}}$ of the interaction is
 \begin{equation}
 f_{\bf{q}} = \frac{g_{c}^2}{ | {\bf{q}} |^{\eta} } e^{ - | {\bf{q}} | /q_c }
 \label{eq:fqdef}
 \; .
 \end{equation}
Here $g_{c}^2$ is some  coupling constant and
$q_{c} \ll k_F$ is an ultraviolet cutoff. 
Interactions of this type have been studied by
Bares and Wen\cite{Bares93}, who showed that
for $\eta = 2 (d-1)$ 
the quasi-particle
residue vanishes logarithmically at the
Fermi surface. The single-particle Green's function
exhibits then anomalous scaling, similar to the one-dimensional
Tomonaga-Luttinger model. 
In the regime $\eta < 2 (d-1)$ the quasi-particle residue remains finite, and
for $\eta > 2 ( d-1 )$ the singularities are even stronger than logarithmic.
Note that  
$\eta \geq d$  corresponds to an interaction 
$ V ( {\bf{r}} - {\bf{r}}^{\prime} )$ that
diverges
for $ | {\bf{r}} - {\bf{r}}^{\prime} | \rightarrow \infty$.
For simplicity, we shall therefore restrict ourselves in this
work to $\eta < d$\cite{Thecase}. 
At the first sight long-range interactions of this type seem
to be irrelevant to real physical systems. 
Note, however, that strong correlations between
electrons can in some cases be described
in terms of long-range gauge forces. 
In fact, recently several authors\cite{Lee89} have used gauge field theories
to explain the experimentally observed non-Fermi liquid
behavior in the normal state of the
high-temperature superconductors\cite{Anderson97}.
As already
pointed out by Bares and Wen\cite{Bares93}, 
the careful analysis of the long-range interactions
of the type (\ref{eq:fqdef}) might also shed
some light onto the nature of the Luttinger liquid
state due the coupling to gauge fields in $d=2$.
Our approach is complementary to the
philosophy adopted in Refs.\cite{Yin96,Chakravarty97},
where a model Green's function of the Luttinger liquid type
in $d > 1$
is assumed without microscopic derivation.
In this paper we shall show
that not all of the 
well known properties of one-dimensional
Luttinger liquids survive in $ d > 1$.
As discussed in Sec.\ref{sec:disorder}, this has
important consequences for the effect of impurities
 on
higher-dimensional Luttinger liquids.

%
%
\section{The single-particle Green's function}
\setcounter{section}{2}
\label{sec:single}

To begin with, let us briefly outline the
calculation of the
single-particle Green's function of 
interacting fermions with dominant forward scattering 
within higher-dimensional bosonization 
\cite{Kopietz97,Luther79,Kopietz95,Kopietz96,Froehlich95}.
First of all, the Fermi surface is subdivided into a finite number
$M$ of small patches with diameter $\Lambda$ such that
$q_c \ll \Lambda \ll k_{F}$. Denoting by ${\bf{k}}^{\alpha}$,
$\alpha = 1, \ldots , M$ a set of wave-vectors on the Fermi surface 
pointing to the centers of the patches, the non-interacting energy dispersion
$\epsilon_{\bf{k}}$ is locally expanded $\epsilon_{ {\bf{k}}^{\alpha} + {\bf{q}} }
- \mu
\approx {\bf{v}}^{ \alpha} \cdot {\bf{q}} +  {\bf{q}}^2/ 2 m^{\alpha}$,
where $\mu = \epsilon_{ {\bf{k}}^{\alpha}}$.  
In the simplest approximation, the inverse effective mass $1/m^{\alpha}$
is set equal to zero. By doing this we neglect the curvature of the Fermi
surface, so
that the patches are completely flat.
Following Refs.\cite{Kopietz97,Kopietz95,Kopietz96,Froehlich95,Lee88}, 
we represent correlation functions as imaginary time
Grassmannian functional integrals
and decouple the two-body interaction by means of 
Hubbard-Stratonovich
fields $V^{\alpha} ( {\bf{q}} , i \omega_m )$
(where $\omega_m = 2 \pi m/{\beta}$ is a bosonic Matsubara frequency, $\beta$ 
is the inverse temperature). 
Integrating over the fermions,
we finally arrive at
the following expression for the real-space imaginary-time Green's function
 \begin{equation}
 G ( {\bf{r} }  -
 {\bf{r}}^{\prime} , \tau - \tau^{\prime} ) 
 = \sum_{\alpha = 1}^{M} e^{i {\bf{k}}^{\alpha} \cdot ( {\bf{r}} - 
{\bf{r}}^{\prime} ) }
 \langle {\cal{G}}^{\alpha} ( {\bf{r}} , {\bf{r}}^{\prime} , \tau , \tau^{\prime} 
)
 \rangle
 \label{eq:Gbos}
 \; .
 \end{equation}
Here $\langle \ldots \rangle $ denotes averaging with respect to the effective 
action
of the Hubbard-Stratonovich field, and ${\cal{G}}^{\alpha}$ is the solution of
 \begin{eqnarray}
 \left[ - \partial_{\tau}  + i {\bf{v}}^{\alpha} \cdot \nabla_{\bf{r}}
 + \frac{ \nabla_{\bf{r}}^2 }{2 m^{\alpha}} - V^{\alpha} ( {\bf{r}} , \tau )
 \right] 
  {\cal{G}}^{\alpha} ( {\bf{r}} , {\bf{r}}^{\prime} , \tau , \tau^{\prime} )
 & = & 
 \nonumber
 \\
& &  \hspace{-40mm}
\delta ( {\bf{r}} - {\bf{r}}^{\prime} ) \delta^{\ast} ( \tau - \tau^{\prime} )
 \; ,
 \label{eq:Gdif}
 \end{eqnarray}
where $V^{\alpha} ( {\bf{r}} , \tau )
= \sum_{ {\bf{q}} \omega_m }
e^{ i ( {\bf{q}} \cdot {\bf{r}} - \omega_m \tau )}
V^{\alpha} ( {\bf{q}} , i \omega_m )$,
and $\delta^{\ast} ( \tau
- \tau^{\prime})$ is the anti-periodic imaginary-time $\delta$-function.
In the limit $1/m^{\alpha} \rightarrow 0$ (i.e.
for linearized energy dispersion) Eq.(\ref{eq:Gdif}) can be solved 
exactly\cite{Schwinger62,Kopietz97,Kopietz95,Lee88}, with the result
 \begin{equation}
  {\cal{G}}^{\alpha} ( {\bf{r}} , {\bf{r}}^{\prime} , \tau , \tau^{\prime} )
 =
  {{G}}^{\alpha}_0 ( {\bf{r}} - {\bf{r}}^{\prime} , \tau - \tau^{\prime} )
 e^{ \Phi^{\alpha} ( {\bf{r}} , \tau ) - \Phi^{\alpha}
 ( {\bf{r}}^{\prime} , \tau^{\prime} ) }
 \; ,
 \label{eq:Schwinger}
 \end{equation}
where $G_{0}^{\alpha} ( {\bf{r}} , \tau )$ 
is the solution of Eq.(\ref{eq:Gdif})
for $V^{\alpha} = 0$, and
 \begin{equation}
 \Phi^{\alpha} ( {\bf{r}} , \tau )
 = \sum_{ {\bf{q}} \omega_m }
 \frac{ e^{ i ( {\bf{q}} \cdot {\bf{r}} - \omega_m \tau )}}{ 
  i \omega_m - {\bf{v}}^{\alpha} \cdot {\bf{q}} } V^{\alpha} ( {\bf{q}} ,
 i \omega_m )
 \label{eq:Phisol}
 \; .
 \end{equation}
Furthermore, for $1/m^{\alpha} \rightarrow 0$ the closed loop 
theorem\cite{Metzner97,Kopietz97,Dzyaloshinskii73}
justifies the Gaussian approximation for the averaging in 
Eq.(\ref{eq:Gbos}), so that we obtain
 \begin{equation}
 G ( {\bf{r} }  , \tau ) 
 = \sum_{\alpha = 1}^{M} 
 e^{i {\bf{k}}^{\alpha} \cdot  {\bf{r}}  }
 G^{\alpha}_0 ( {\bf{r} }  , \tau  ) 
 e^{Q^{\alpha \alpha} ( {\bf{r}} , \tau )}
 \; ,
 \label{eq:Galphasum}
 \end{equation}
where for later reference we have introduced a 
Debye-Waller factor
 $Q^{\alpha \alpha^{\prime}} ( {\bf{r}} , \tau )$
with two patch indices $\alpha$ and $\alpha^{\prime}$,
 \begin{equation}
 Q^{\alpha \alpha^{\prime} } ( {\bf{r}} , \tau ) = S^{\alpha \alpha'} (0, 0) 
 - S^{\alpha \alpha'} ( {\bf{r}} , \tau 
)
 \; ,
 \label{eq:Deb}
 \end{equation}
 \begin{equation}
 S^{\alpha \alpha'} ( {\bf{r}} , \tau )
 = \frac{1}{ {\cal{\beta V}} } \sum_{ {\bf{q}} \omega_m }
 \frac{ f^{\rm RPA} ( {\bf{q}} , i \omega_m ) \cos
 ( {\bf{q}} \cdot {\bf{r}} - \omega_m \tau ) }{ [ i \omega_m -
 {\bf{v}}^{\alpha} \cdot {\bf{q}}  ] [ i \omega_m -
 {\bf{v}}^{\alpha'} \cdot {\bf{q}}  ] }
 \; .
 \label{eq:Sdef}
 \end{equation}
Here ${\cal{V}}$ is the volume of the system, and
 \begin{equation}
 f^{\rm RPA} ( {\bf{q}} , i \omega_m )
 = \frac{f_{\bf{q}} }{  1 + f_{\bf{q}} \Pi_0 ( {\bf{q}} , i \omega_m ) }
 \label{eq:frpadef}
 \end{equation} 
is the screened interaction within the random-phase approximation (RPA).

It should be mentioned that a priori it is not clear whether 
it is allowed to linearize the energy dispersion
(corresponding to taking the limit $1/m^{\alpha} \rightarrow 0$)
or not. As discussed in Ref.\cite{Kopietz97},
in dimensions $d >1$ this approximation becomes 
more questionable than in $d=1$, because for momenta ${\bf{q}}$
parallel
to the Fermi surface the linear term ${\bf{v}}^{\alpha} \cdot {\bf{q}}$
vanishes, so that the quadratic term ${\bf{q}}^2 / 2 m^{\alpha}$
is the leading one.
In Refs.\cite{Kopietz97,Kopietz96} a systematic method
for including into the
bosonization procedure 
the effects associated with the
finiteness of $1/m^{\alpha}$ 
has been developed.
Although the expressions derived in these works have
so far not been analyzed in detail, 
it can be shown\cite{Kopietz97} that for the calculation of the
momentum distribution in the presence of the singular
density-density interactions discussed here it is indeed
sufficient to work with linearized energy dispersion.

\begin{figure}
\begin{center}
\epsfysize4.5cm 
\epsfbox{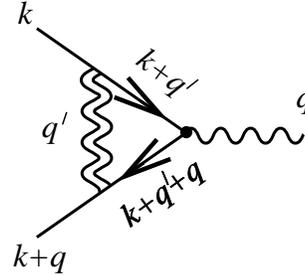}
\vspace{0.4cm}
\caption{
Leading interaction correction to the
density vertex. Here, $k$ is short for $({\bf{k}},i\tilde{\omega}_n )$ 
and $q$ is short for $({\bf{q}},i{\omega}_m )$. The double wavy line is
 the RPA screened interaction, the wavy line is the external
field, and the
solid arrows denote non-interacting single-particle
Green's functions.
}
\label{fig:vertex}
\end{center}
\end{figure}

If the system is a Fermi liquid, then
the quasi-particle residue for wave-vectors close to
${\bf{k}}^{\alpha}$ can be identified with \cite{Kopietz97,Kopietz95}
 \begin{equation}
 Z^{\alpha} = e^{ S^{\alpha \alpha} ( 0 , 0 ) }
 \label{eq:Z}
 \; .
 \end{equation}
On the other hand, non-Fermi liquid behavior 
in the single-particle Green's function
manifests itself in the divergence of 
$S^{\alpha \alpha} ( 0 , 0 )$ in the limit ${\cal{V}} \rightarrow \infty$ and
$\beta \rightarrow \infty$. In fact, this quantity is closely related
to the leading non-trivial vertex correction 
shown in Fig.\ref{fig:vertex},
 \begin{eqnarray}
 \Lambda_1 ( {\bf{k}} , i \tilde{\omega}_{n} ; {\bf{q}} , i \omega_m )
 & = &  - \frac{1}{ {\beta \cal{V}} } \sum_{ {\bf{q}}^{\prime}  \omega_{{m}^{\prime}} }
 f^{\rm RPA} ( {\bf{q}}^{\prime} , i \omega_{{m}^{\prime}} )
 \nonumber
 \\
 & & \hspace{-32mm} \times
 G_0 ( {\bf{k}} + {\bf{q}}^{\prime}, i \tilde{\omega}_{n + m^{\prime}} )
 G_0 ( {\bf{k}} + {\bf{q}}^{\prime} + {\bf{q}} , 
 i \tilde{\omega}_{n + m^{\prime} + m} )
 \; ,
 \label{eq:vertex1}
 \end{eqnarray} 
where $\tilde{\omega}_{n} = 2\pi ( n + \frac{1}{2} ) /{\beta}$, and
$G_{0} ( {\bf{k}} , i \tilde{\omega}_n )
= [ i \tilde{\omega}_n - \epsilon_{\bf{k}} + \mu ]^{-1}$ 
is the non-interacting Matsubara Green's function.
Noting that for linearized energy dispersion
 \begin{equation}
 G_0 ( {\bf{k}}^{\alpha} + {\bf{q}} , i \tilde{\omega}_n ) 
 = \frac{1}{ i \tilde{\omega}_n - {\bf{v}}^{\alpha} \cdot {\bf{q}} }
 \label{eq:G0def}
 \; ,
 \end{equation}
and comparing Eqs.(\ref{eq:vertex1}) and (\ref{eq:Sdef}), it is clear that 
in the limit $\beta \rightarrow \infty$
we may identify\cite{Incase}
 \begin{equation}
 S^{\alpha \alpha} ( 0 , 0 ) = - \Lambda_1 ( {\bf{k}}^{\alpha} , 0 ; 0 , 0 )
 \equiv - \Lambda_1 ( 0 )
 \; .
 \label{eq:SLambda}
 \end{equation}
Thus, Eq.(\ref{eq:Z}) amounts to an exponentiation of the first
non-trivial vertex correction.

Obviously, in the above derivation we have ignored 
all scattering processes that transfer momenta between
different patches on the Fermi surface
(the so-called around-the-corner processes).  
Only then the closed loop
theorem\cite{Metzner97,Kopietz97,Dzyaloshinskii73}
is valid, which implies  that
the effective interaction in Eq.(\ref{eq:Sdef})
is simply given by the RPA. 
The neglect of the around-the-corner processes
is one of the main approximations 
inherent in higher-dimensional bosonization\cite{Kopietz97,Luther79}.
We are not awave of any quantitative calculation
of the single-particle Green's function which 
takes these processes explicitly into account.
However, for $q_c \ll \Lambda$ it is
reasonable to expect that these processes can indeed
be neglected, provided the 
global topology of the Fermi surface does not play a crucial role.
The reason is that for $q_c \ll \Lambda$
the number of around-the-corner scattering 
processes is always paramatrically smaller than
the number of forward scattering processes\cite{Kopietz97}.

Let us now evaluate the above expressions
in the case of the singular interaction
of the type (\ref{eq:fqdef}).
Previously, Bares and Wen\cite{Bares93}
have considered only static properties,
such as the momentum distribution. 
To the best of our knowledge, a calculation of dynamic properties, such as the
frequency-dependent density of states $\rho ( \omega )$,
cannot be found in the literature. 
Such a calculation is straightforward within the framework of higher-dimensional
bosonization. The calculations are simplified if one uses the fact that
the non-Fermi liquid  behavior is due to the existence of
a collective plasmon mode $\omega_{\bf{q}}$ in the density-density correlation
function. The long-wavelength dispersion of this mode is 
\cite{Kopietz97,Castellani97}
 \begin{equation}
 \omega_{\bf{q}} \sim \frac{ v_F \kappa}{\sqrt{d}} \left( \frac{ | {\bf{q}} |  
}{\kappa} 
 \right)^{1- \eta /2 }
 \label{eq:plasmon}
 \; ,
 \end{equation}
where $\kappa = ( g_{c}^2 \nu )^{1 / \eta}$, and $\nu$
is the non-interacting density of states. 
It turns out that this mode completely determines the
infrared behavior of the single particle Green's function.
Because for small wave-vectors $\omega_{\bf{q}} \gg v_{F}  | {\bf{q}} |$,
we may 
restrict the frequency integration in Eq.(\ref{eq:Sdef}) to the regime
$ | \omega_m | \gg v_{F} | {\bf{q}} | $. Evidently, in this regime  
the terms ${\bf{v}}^{\alpha} \cdot {\bf{q}}$ in the denominator
of Eq.(\ref{eq:Sdef}) can be 
ignored, which implies that the {\it{leading 
long-distance or long-time behavior of the Debye-Waller
factor 
$Q^{\alpha \alpha^{\prime} } ( {\bf{r}} , \tau )$ 
does not
depend on the patch indices $ \alpha$ and $\alpha^{\prime}$.}}
Furthermore, the RPA interaction
$f^{\rm RPA} ( {\bf{q}} , i \omega_m )$  
can be replaced by its
limit for $ | \omega_m | \gg v_F | {\bf{q}}|$, which is given by
 \begin{equation}
 f^{\rm RPA} ( {\bf{q}} , i \omega_m ) \approx
 \frac{1}{\nu} \left( \frac{\kappa }{ | \bf{q} | }\right)^{\eta}
 \frac{  \omega_m^2}{ \omega_m^2 
 + \omega_{\bf{q}}^2 }
 e^{-|{\bf q}|/q_{c}}
 \; .
 \label{eq:frpaapprox}
 \end{equation}
After some straightforward manipulations we find that
in the limit ${\cal{V}} \rightarrow \infty$ and
$\beta \rightarrow \infty$ the dominant contribution to the
Debye-Waller-factor
can be written as
 \begin{eqnarray}
 Q^{\alpha \alpha'} ( {\bf{r}} , \tau )
& \approx & -\frac{\sqrt{d}}{2}
 \frac{\kappa^{\frac{\eta}{2}}}{k_{F}^{d-1}} \frac{\Omega_{d-1}}
 {\Omega_{d}} \int_{0}^{\pi} d \vartheta \sin^{d-2} \vartheta 
 \nonumber \\
 & & \hspace{-21mm}
 \times  \int_{0}^{\infty} d  q\; q^{d-2-\frac{\eta}{2}}  e^{-\frac{q}{2 q_c}} 
 \left[ 1- e^{-\omega_{\bf{q}} | \tau | } \cos \left(q 
 | {\bf{r}} |  \cos \vartheta \right) \right]
 \; .
 \label{eq:Qaa}
\end{eqnarray}
Here 
$\Omega_d$ is the surface area of a $d$-dimensional unit sphere, and
we have used $\nu = (\Omega_{d}/(2\pi)^d) m k_F^{d-2}$.

Let us now focus on the
case $\eta 
= 2(d-1)$. As mentioned above, in this case 
perturbation theory is plagued by logarithmic singularities. 
The long-distance behavior of $Q^{\alpha \alpha'}
\left({\bf{r}},0 \right)$ can be easily extracted, and we obtain
for $ | {\bf{r}} | \rightarrow \infty $
\begin{equation}
e^{Q^{\alpha \alpha'}({\bf{r}},0)} \sim \left({q_{c} |{\bf{r}}|}
\right)^{- \gamma} \; ,
\end{equation}
where the anomalous dimension is given by
\begin{equation}
\gamma = \frac{\sqrt{d}}{2} \left(\frac{\kappa}{k_F}\right)^{d-1} = \left(
\frac{d m g_{c}^{2}}{2^{d+1} \pi^{\frac{d}{2}} \Gamma (\frac{d}{2}) k_{F}^{d}}
\right)^{\frac{1}{2}} \; .
\label{eq:gamma}
\end{equation}
From this we can obtain the momentum distribution 
$n^{\alpha} ( {\bf{q}} ) = n ( {\bf{k}}^{\alpha} + {\bf{q}} )$
near the non-interacting Fermi
surface. For $\gamma < 1$ we find
\begin{equation}
n^{\alpha}({\bf{q}}) = \frac{1}{2} \left[1-{\rm{sgn}} (q_{\parallel}^{\alpha})
\Gamma
(1-\gamma) \frac{\sin(\frac{\pi}{2} \gamma)}{\frac{\pi}{2} \gamma}
\left|\frac{q_{\parallel}^{\alpha}}{q_{c}}
\right|^{\gamma} \right]\; ,
\end{equation}
where $q^{\alpha}_{\parallel}$ is the projection of ${\bf{q}}$ onto 
${\bf{k}}^{\alpha}$. 
Similarly we can compute $Q^{\alpha \alpha'}(0,\tau)$
for $\eta = 2 ( d - 1)$.
We find for $\tau \rightarrow \infty$
\begin{equation}
 e^{Q^{\alpha \alpha'}(0,\tau)} \sim \left(\frac{\kappa}{q_{c}}\right)^{\gamma}
\left(\kappa {v_{F} |\tau |} \right)^{- \tilde{\gamma}}\; .
\end{equation}
where
for $1 \leq d < 2$ the {\it{dynamic anomalous dimension}} $\tilde{\gamma}$ is
\begin{equation}
\tilde{\gamma} = \frac{\gamma}{2-d} \quad \; .
\label{eq:tildegamma} 
\end{equation}
Recall that we have excluded the case $d=2$, which for
$\eta = 2 (d-1)$ corresponds to $\eta =2$\cite{Thecase}. 
The small frequency behavior of the density of states can now be calculated,
and we finally obtain
\begin{equation}
\rho(\omega) = -\frac{1}{\pi}  \rm{Im} G
({\bf{r}}=0,\omega + i 0^{+} ) \propto |\omega|^{\tilde{\gamma}}\; .
\label{eq:DOSano} 
\end{equation}
We conclude that in dimensions $1 < d < 2$
the singularities in the momentum distribution and
in the density of states of our higher-dimensional Luttinger liquid
are characterized by {\it{different anomalous dimensions}}.
Although we have shown this
only for a specific class of models, we believe 
that this is a general property of higher-dimensional Luttinger liquids. 
The reason is that in $d > 1$ the Fermi surface consists of a $d-1$-dimensional
continuum, so that loop integrations will always involve 
some kind of angular averaging over the Fermi surface. This
destroys the symmetry between space and time variables, which is 
only preserved in $d=1$.

%
%
\section{The polarization} 
\setcounter{section}{3}
\label{sec:pol}
Let us now study how singular interactions of the type
(\ref{eq:fqdef}) affect the low-frequency behavior of
the irreducible polarization $\Pi_{\star} ( {\bf{q}} , \omega )$.
First of all, let us recall that
for small
momenta ${\bf{q}}$ and small frequencies $\omega$
the leading singular corrections to
$\Pi_{\star} ( {\bf{q}} , \omega )$ 
cancel, so that
 \begin{equation}
 \Pi_{\star} ( {\bf{q}} , \omega ) \approx \Pi_{0} ( {\bf{q}} , \omega )
 \label{eq:Piir}
 \; \; , \; \;
\mbox{$ | {\bf{q}} | \ll k_F \; , \; |\omega | \ll E_F$}
 \; ,
 \end{equation}
where $\Pi_0$ is the non-interacting polarization.
Eq.(\ref{eq:Piir}) is a consequence of a complete
cancellation of the most singular
self-energy and vertex corrections, which happens not only in
$d=1$\cite{Dzyaloshinskii73} but also
in higher dimensions\cite{Metzner97,Kopietz97,Kopietz95}.
For $ | {\bf{q}} | = O ( k_F)$ this cancellation does in general not occur.
In fact, in the one-dimensional Tomonaga-Luttinger model it is well known
that for momentum transfers $ | {\bf{q}} | \approx 2 k_F$ 
the irreducible polarization (as well as various other two-particle
correlation functions) exhibits anomalous scaling behavior, 
very much like the
single particle Green's function, although with different  anomalous 
exponents\cite{Solyom79}.
We now show that this property
of one-dimensional Luttinger liquids does in general not survive
in higher dimensions.
We would like to emphasize that the interaction of the type 
(\ref{eq:fqdef}) has no momentum transfers of order $O ( k_F)$, so
that bosonization can be used to sum the dominant singularities. It is
the {\it external} momentum transfer of the irreducible polarization
which is approximately $2 k_F$.

Within the same approximations as described in Sec.\ref{sec:single},
the real-space, imaginary-time Fourier transform
of  
$\Pi_{\star} ( {\bf{q}} , i \omega_m )$ can be written as
 \begin{eqnarray}
 \Pi_{\star} ( {\bf{r}} - {\bf{r}}^{\prime} , \tau - \tau^{\prime} )
 & = & - \sum_{\alpha \alpha' = 1}^{M} 
 e^{ i ( {\bf{k}}^{\alpha} - {\bf{k}}^{\alpha'} ) 
 \cdot ( {\bf{r}} - {\bf{r}}^{\prime} ) }
 \nonumber
 \\
 & \times &
 \langle 
 {\cal{G}}^{\alpha} ( {\bf{r}} , {\bf{r}}^{\prime} , \tau , \tau^{\prime} )
 {\cal{G}}^{\alpha'} ( {\bf{r}}^{\prime} , {\bf{r}} , \tau^{\prime} , \tau )
 \rangle
 \label{eq:Pi1}
 \; .
 \end{eqnarray}
Performing the averaging within the Gaussian approximation, we obtain
 \begin{eqnarray}
 \Pi_{\star} ( {\bf{r}} , \tau )
 & = & - \sum_{\alpha \alpha' = 1}^{M}
 e^{ i ( {\bf{k}}^{\alpha} - {\bf{k}}^{\alpha'} ) 
 \cdot  {\bf{r}}  }
 {{G}}^{\alpha}_0 ( {\bf{r}} , \tau  )
 {{G}}^{\alpha'}_0 ( - {\bf{r}} , - \tau )
 \nonumber
 \\
 & \times &
 \exp \big[  Q^{\alpha \alpha}  ( {\bf{r}} , \tau ) 
  + Q^{\alpha' \alpha'} ( {\bf{r}} , \tau ) \big. \nonumber \\ 
 &  & \hspace{3mm} \big. -  Q^{\alpha \alpha'}  ( {\bf{r}} , \tau ) 
  - Q^{\alpha' \alpha} ( {\bf{r}} , \tau )  \big]
 \label{eq:Pi2}
 \; .
 \end{eqnarray}
For $| {\bf{q}} | \ll k_F$ the Fourier transform
$\Pi_{\star} ( {\bf{q}} , i \omega_m )$ of
Eq.(\ref{eq:Pi2}) 
is dominated by  the diagonal terms
${\bf{k}}^{\alpha} = {\bf{k}}^{\alpha'}$, and we recover
Eq.(\ref{eq:Piir}). Similarly, for $ | {\bf{q}} - 2 {\bf{k}}^{\alpha} |
\ll k_F$, Eq.(\ref{eq:Pi2}) is
dominated by ${\bf{k}}^{\alpha} = - {\bf{k}}^{\alpha'}$, so that
the Fourier transform of
$\Pi_{\star} ( 2 {\bf{k}}^{\alpha} + {\bf{q}} , i \omega_m )$ is for
$ | {\bf{q}}  | \ll k_F$ given by
 \begin{equation}
 \Pi_{\star}^{2 {\bf{k}}^{\alpha} } ( {\bf{r}} , \tau ) =
 \Pi_{0}^{ 2 {\bf{k}}^{\alpha} } ( {\bf{r}} , \tau )
 e^{ {Q}^{2 \bf{k}^{\alpha}} ( {\bf{r}} , \tau ) }
 \label{eq:Pi2kF}
 \; ,
 \end{equation}
where (using the notation ${\bf{k}}^{\bar{\alpha}} = - {\bf{k}}^{\alpha}$)
 \begin{equation}
 \Pi_{0}^{ 2 {\bf{k}}^{\alpha} } ( {\bf{r}} , \tau )
 = - e^{ i  2 {\bf{k}}^{\alpha} \cdot {\bf{r}} }
 {{G}}^{\alpha}_0 ( {\bf{r}} , \tau  )
 {{G}}^{\bar{\alpha}}_0 ( - {\bf{r}} , - \tau )
 \label{eq:Pi02kF} 
 \; ,
 \end{equation}
and the Debye-Waller factor is given by
 \begin{eqnarray}
 & & \hspace{-8mm} {Q}^{2 \bf{k}^{\alpha}} ( {\bf{r}} , \tau )  
 \nonumber
 \\
 & & \hspace{-8mm} =
 Q^{\alpha \alpha}
 ( {\bf{r}} , \tau ) + Q^{\bar{\alpha} \bar{\alpha}} ( {\bf{r}} , \tau )
  -  Q^{\bar{\alpha} \alpha} ( {\bf{r}} , \tau )
 - Q^{\alpha \bar{\alpha}} ( {\bf{r}} , \tau )
 \nonumber
 \\
 &  & \hspace{-8mm} = 
  \frac{1}{ {\beta \cal{V}} } \sum_{ {\bf{q}} \omega_m }
 f^{\rm RPA} ( {\bf{q}} , i \omega_m )  
 \frac{ 4 ( {\bf{v}}^{\alpha} \cdot {\bf{q}} )^2 }{
 \left[ \omega_m^2 + ( {\bf{v}}^{\alpha} \cdot {\bf{q}} )^2 \right]^2}
\nonumber
 \\
&  & \hspace{-3mm} \times
[ 1 - \cos
 ( {\bf{q}} \cdot {\bf{r}} - \omega_m \tau ) ]
 \; .
 \label{eq:Qbardef}
 \end{eqnarray}
Note that this Debye-Waller factor
is different from the Debye-Waller factor
$Q^{\alpha \alpha} ( {\bf{r}} , \tau )$  
of the single-particle Green's function.
It should also be mentioned that for any finite number
of flat patches the non-interacting polarization
$\Pi_0^{2 {\bf{k}}^{\alpha} } ( {\bf{r}} , \tau )$ exhibits
in $d > 1$
logarithmic singularities due to the
artificial nesting symmetry, which is generated by replacing the
curved Fermi surface by a finite number of flat patches.
However, we have isolated 
possible non-trivial effects due to
interactions in the Debye-Waller factor
 ${Q}^{2 \bf{k}^{\alpha}} ( {\bf{r}} , \tau ) $.

It is not difficult to show that in one dimension
Eq.(\ref{eq:Qbardef})  correctly reproduces 
the irreducible polarization
of the Tomonaga-Luttinger model for momenta
close to $2 k_F$\cite{Solyom79}.
Note that in this case the large-distance and long-time
decay of $\Pi_{\star} ( {\bf{r}} , \tau )$ is characterized by non-universal power laws,
analogous to the single-particle Green's function. 
The anomalous scaling manifests itself via a logarithmic growth of the
Debye-Waller factor (\ref{eq:Qbardef}) in the large-distance or long-time
limit.  
The important observation is now that
in the higher-dimensional non-Fermi liquid  state 
due to the
singular density-density interactions 
(\ref{eq:fqdef}) the irreducible polarization
 $\Pi_{\star}^{2 {\bf{k}}^{\alpha} } ( {\bf{r}} , \tau )$ 
{\it{does not exhibit any anomalous scaling}}.
The case of the Tomonaga-Luttinger model
(corresponding to $d=1$ and $\eta = 0$) is special, because
in this case the collective plasmon mode is linear
in $| {\bf{q}}|$, so that it has the same order of magnitude as the
single-particle excitation energy of the bare fermions.
The fact that in $d> 1$ the leading singularities of
$Q^{\alpha \alpha^{\prime} } ( {\bf{r}} , \tau )$
cancel in ${Q}^{2 {\bf{k}}^{\alpha}} ( {\bf{r}} , \tau )$ 
follows trivially
from the observation that the infrared behavior of the Debye-Waller factor
 $Q^{\alpha \alpha'} ( {\bf{r}} , \tau )$ is
independent of the patch indices, 
so that the four terms in the second line of Eq.(\ref{eq:Qbardef}) cancel. 
As already mentioned in Sec.\ref{sec:single}, this is due to the fact
that
for $\eta > 0$ the plasmon energy at long wavelengths is larger
than $ {\bf{v}}^{\alpha} \cdot {\bf{q}}$, so that
the integrand in Eq.(\ref{eq:Qbardef}) contains an additional small
factor of $ ( {\bf{v}}^{\alpha} \cdot {\bf{q}} / \omega_{\bf{q}} )^2 $
as compared with the integrand in Eq.(\ref{eq:Sdef}).
This factor cancels the leading singularity.

\begin{figure}
\begin{center}
\epsfysize2.3cm 
\epsfbox{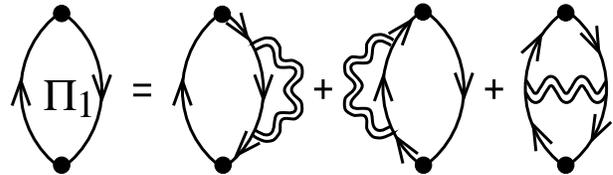}
\vspace{4mm}
\caption{
Leading interaction corrections to the
irreducible polarization for momentum transfers close to
$2  {\bf{k}}^{\alpha}$.
The first two diagrams correspond to self-energy corrections (see Eq. 
(\ref{eq:Pi1self})),
the third diagram  is the leading vertex correction (see Eq. 
(\ref{eq:Pi1vert})).
}
\label{fig:pol}
\end{center}
\end{figure}

It is instructive to relate the constant part
of the Debye-Waller factor (\ref{eq:Qbardef}) 
to vertex corrections, similar
to Eq.(\ref{eq:SLambda}).
Consider therefore the leading interaction corrections to the irreducible 
polarization for momentum transfers close to
$2 {\bf{k}}^{\alpha}$ shown in Fig.\ref{fig:pol}.
Calculating these corrections in real space and imaginary time,
we find that they can be expressed
in terms of the Debye-Waller factor
$Q^{\alpha \alpha^{\prime}} ( {\bf{r}} , \tau )$
defined in Eqs.(\ref{eq:Deb}) and (\ref{eq:Sdef})
as follows,
\begin{eqnarray}
 \Pi^{\alpha \alpha}_{1,\rm{self}} ({\bf{r}},\tau) & = & 2
 \Pi^{\alpha \alpha}_{0} ({\bf{r}},\tau)  Q^{\alpha \alpha} ({\bf{r}},\tau)
 \; ,
 \label{eq:Pi1self}
 \\
 \Pi^{\alpha \alpha'}_{1,\rm{vertex}}({\bf{r}},\tau)  & = & -2 
 \Pi^{\alpha \alpha'}_{0} ({\bf{r}},\tau) 
 Q^{\alpha \alpha'} ({\bf{r}},\tau)
\label{eq:Pi1vert}
 \; .
\end{eqnarray}
From these expressions it is obvious that an expansion of Eq.(\ref{eq:Qbardef})
to first order in the RPA-interaction exactly
reproduces perturbation theory. 
Not surprisingly, bosonization amounts to an exponentiation
of the leading perturbative corrections 
in real space and imaginary time.
Just like in the case of the single-particle Green's function
(see Eq.(\ref{eq:SLambda})),
we can identify the constant part
$S^{2\bf{k}^{\alpha}}(0,0)$
of the Debye-Waller factor (\ref{eq:Qbardef}) with
a certain combination of vertex corrections.
Assuming again that
$S^{2\bf{k}^{\alpha}}(0,0)$ is finite, we see that
in the  limit  of large distances or long times interactions
simply lead to an overall multiplicative renormalization
of the non-interacting polarization
by a factor of
$e^{S^{2\bf{k}^{\alpha}}(0,0)}$.
Note that
$S^{2\bf{k}^{\alpha}}(0,0)$ 
is always greater 
than zero and is finite for interactions of the type  (\ref{eq:fqdef}) 
in dimensions $d > 1$.
From Eq.(\ref{eq:vertex1}) it is easy to see that
\begin{eqnarray}
S^{\alpha \alpha'} (0,0) &=&  -\Lambda_1
({\bf{k}}^{\alpha'},0;{\bf{k}}^{\alpha} - {\bf{k}}^{\alpha'},0)
\nonumber \\
 &\equiv& -\Lambda_1
 (|{\bf{k}}^{\alpha} - {\bf{k}}^{\alpha'}|)\; . 
\end{eqnarray}
It follows that
$S^{2{\bf{k}}^{\alpha}}(0,0)$ can also be written as
\begin{equation}
 S^{2\bf{k}^{\alpha}}(0,0) = 2[\Lambda_1(2 k_F) - \Lambda_1(0)]\; ,
 \label{eq:Svertexlambda}
\end{equation}
which should be compared with Eq.(\ref{eq:SLambda}).
From Eq.(\ref{eq:Svertexlambda}) we can
obtain an alternative explanation why
for $\eta \ge 2 (d-1)$ the non-Fermi liquid 
nature of the single-particle Green's function
does not affect the polarization: although in this case
both vertex corrections
$\Lambda_1 ( 2 k_F )$ and $\Lambda_1 (0)$ are divergent,
{\it{their difference 
$\Lambda_1 ( 2 k_F ) - \Lambda_1 (0)$ 
remains finite}}.
In this respect our higher-dimensional non-Fermi  liquid
behaves fundamentally different than the one-dimensional
Tomonaga-Luttinger model, where 
$\Lambda_1 ( 2 k_F )$, $\Lambda_1 (0)$, {\it{and
the difference}}
$\Lambda_1 ( 2 k_F ) - \Lambda_1 (0)$ 
are all singular.

%
%
\section{Disordered Luttinger liquids in higher dimensions}
\label{sec:disorder}

The conventional strategy to study interactions
in disordered Fermi systems
is to treat first the effect of disorder as accurate as possible, 
and only after that incorporate the effects of interactions perturbatively.
It has recently been pointed out by several authors\cite{Chakravarty97,Varma97}
that for strongly interacting Fermi systems this approach may not correctly
capture the interplay between disorder and interactions in low dimensions.
In fact, in some cases it may be better to 
solve first the interaction problem without disorder, and then study
the additional effect of disorder.
Here we would like to point out that in this approach it is crucial
to distinguish between the single-particle scattering time
$\tau_{1} ( \omega )$ (which is defined in terms of the
single-particle Green's function) and the transport
scattering time $\tau_{\rm tr} ( \omega )$ (which is related to the conductivity).
This difference is important {\it{even if the impurity potential
is dominated by $s$-wave scattering.}}
The reason is that in non-Fermi liquids vertex corrections play a very important role,
so that even for a $\delta$-function correlated random potential
the low-frequency behavior of $\tau_{1} ( \omega )$ and $\tau_{\rm tr} ( \omega )$ can be
completely different. 

To substantiate this point, let us use the
above non-perturbative results of Sec.\ref{sec:single} and 
\ref{sec:pol} to study
the effect of disorder on the clean system
perturbatively.
As usual we model the disorder
by coupling the charge density to a 
Gaussian random potential $U ( {\bf{r}} )$ with zero average
and $\delta$-function correlations
 \begin{equation}
 \overline{ U ( {\bf{r}} ) U ( {\bf{r}}^{\prime} ) } = \gamma_{0}
  \delta ( {\bf{r}} - {\bf{r}}^{\prime} )
 \; ,
 \label{eq:random}
 \end{equation}
where the overline denotes averaging over the probability 
distribution of $U ( {\bf{r}} )$ and the parameter $\gamma_{0}$
is a measure of the strength of the disorder.
Within the lowest order Born approximation
the disorder gives rise to the following self-energy correction 
to the single-particle Green's function,
 \begin{equation}
 \Sigma_1 ( \omega )  =  \frac{\gamma_{0}}{\cal{V}} \sum_{\bf{k}} 
 G ( {\bf{k}} , \omega  )
  =  \gamma_{0} G ( {\bf{r}} = 0 , \omega )
 \; ,
 \label{eq:Born}
 \end{equation}
where $G$ is the Green's function of the interacting
many-body system without disorder.
The single-particle scattering time $\tau_1 ( \omega )$ is 
defined by 
 \begin{equation}
 {  1/ \tau_1 ( \omega ) }
 = - 2 {\rm Im} \Sigma_{1} ( \omega + i 0^{+} )
 \; .
 \label{eq:tau1def}
 \end{equation}
Using Eq.(\ref{eq:DOSano}), we see that 
in our model for $\eta = 2 ( d-1 )$
the single-particle scattering time diverges within lowest order
Born approximation as $\tau_1 ( \omega ) \propto | \omega |^{- \tilde{\gamma} }$,
where the dynamic anomalous dimension
$\tilde{\gamma}$ is given in Eq.(\ref{eq:tildegamma}).
As shown in Ref.\cite{Chakravarty97}, this result turns out to be
only self-consistent for
$\tilde{\gamma} > 1/2$. 
For smaller $\tilde{\gamma}$ this divergence is removed
if one uses instead of Eq.(\ref{eq:Born})
the {\it{self-consistent}} Born approximation,
which
amounts to the replacement
 \begin{equation}
 G ( {\bf{k}} , \omega )  \rightarrow 
 [ G^{-1} ( {\bf{k}} , \omega ) - \Sigma_1 ( \omega ) ]^{-1}
\label{eq:replacement}
 \end{equation}
on the right-hand side of Eq.(\ref{eq:Born}).
However, even if $\tau_1 ( \omega )$ diverges,
it would be
{\it{incorrect to conclude that the conductivity
and the diffusion coefficient diverge as well}}.
Such a conclusion
is only valid as long as vertex
corrections are unimportant, 
which is certainly not the case for higher-dimensional Luttinger liquids.
To see this more clearly, one should keep in mind that  
the Drude formula
$\sigma = n e^2 \tau_{\rm tr} / m^{\ast}$ for the conductivity
($n$ is the density of the electrons and $m^{\ast}$ is their effective mass)
involves the {\it{transport scattering time}} $\tau_{\rm tr}$, which
for interacting fermions is in general not identical
with the single-particle scattering time
$\tau_1$ defined in Eq.(\ref{eq:tau1def}).
Microscopically, the transport 
scattering time $\tau_{\rm tr} ( \omega )$ can
be calculated from the imaginary part of the
memory function $M ( \omega + i 0^{+})$\cite{Goetze72,Luther74},
 \begin{equation}
 1/ {\tau_{\rm tr} ( \omega ) } = {\rm Im} M ( \omega + i 0^{+} )
 \; .
 \label{eq:tautrdef}
 \end{equation}
For arbitrary complex frequencies $z$
the memory function is defined in terms of the conductivity via
 \begin{equation}
 \sigma (z ) =  \frac{n e^2}{m^{\ast}} \frac{i}{z + M ( z )}
 \; .
 \end{equation}
Expanding $M ( z)$ to first order in 
the impurity concentration, one obtains in $d$ dimensions\cite{Goetze72}
 \begin{equation}
 M ( z ) = \frac{\gamma_{0}}{d n {\cal{V}} } \sum_{ {\bf{q}} } \frac{ {\bf{q}}^2 }{ m^{\ast} }
 \frac{ \Pi ( {\bf{q}} , z ) - \Pi ( {\bf{q}} , 0 ) }{z}
\label{eq:Mexp}
 \; .
\end{equation}
Thus, to lowest order in $\gamma_{0}$
the transport scattering time is determined by the (reducible) polarization
$\Pi ( {\bf{q}} , z )$
of the interacting many-body system.
Of course, for non-interacting electrons 
and a $\delta$-function correlated random
potential  (i.e. $s$-wave impurity scattering) 
both definitions (\ref{eq:tau1def}) and (\ref{eq:tautrdef}) 
lead to identical results for the scattering times 
in the limit $\omega \rightarrow 0$.
However, for the
class of higher-dimensional Luttinger liquids
studied in this paper  the low-frequency behavior of
$\tau_1 ( \omega )$ and $\tau_{\rm tr} ( \omega )$ is completely
different.
This is obvious from the fact that in these systems
the polarization does not exhibit any
anomalous behavior, so that $\tau_{\rm tr} ( 0 )$ 
is simply a constant.
Thus, to lowest order in the impurity concentration the
static conductivity $\sigma (0)$ is finite.
Note that this implies also a finite diffusion coefficient
${\cal{D}} = \sigma ( 0) / ( e^2 \partial n / \partial \mu )$,
where $\partial n / \partial \mu = \lim_{ {\bf{q}} \rightarrow 0} \Pi_{\star} ( {\bf{q}} , 0 )$
is the compressibility\cite{Finkelstein83}. 
Hence,  within  
a perturbative approach to lowest order in the impurity
concentration there is no sign for a divergent conductivity
in our higher-dimensional Luttinger liquids,
even in the regime where the single-particle scattering time
$\tau_1 ( \omega )$ diverges for $\omega \rightarrow 0$.
Of course, the true low-frequency behavior of the conductivity
most likely requires a non-perturbative treatment of disorder, and
cannot be calculated by simply expanding the memory function
to first order in the impurity concentration.

%
%
\section{Summary}
\setcounter{section}{5}
\label{sec:discussion}

In this work we have used higher-dimensional bosonization to study
the correlation functions of a particular class of higher-dimensional
non-Fermi liquids, which are characterized by singular density-density
interactions diverging in momentum space as $q^{- \eta }$ for small
momentum transfers $q$. We have paid particular attention to the
case $\eta = 2 (d-1)$ in dimensions $1 \leq d < 2$, 
which can be considered as a possible generalization of the
Tomonaga-Luttinger model to higher dimensions.
The main result of this work is the observation that
in $d>1$ not all of the well-known properties of the Tomonaga-Luttinger model
survive. 
Thus, the correlation functions of higher-dimensional Luttinger liquids
cannot be obtained by straightforward generalizations of the
correlation functions of the one-dimensional models.
In particular, we have shown that in $d>1$ the scaling properties of the
single-particle Green's function cannot be characterized by a single anomalous dimension.
Furthermore, unlike in $d=1$, in higher dimensions
the polarization for momentum transfers close to $2k_F$ does
not show any non-Fermi liquid behavior. 
Although we have obtained these results for a rather special class of models, we believe
that they are quite general and reflect generic properties of higher-dimensional
Luttinger liquids.
In the language of many-body perturbation theory,
the qualitative difference between the low-frequency behavior of the
single-particle Green's function and the polarization is 
due to the importance of vertex corrections,
which in $d > 1$ tend to cancel in
gauge invariant correlation functions such as the polarization,
{\it{even for large momentum transfers}}.

We would like to point out that 
our results have been obtained for singular 
{\it{density-density interactions}}.
Recently many authors\cite{Lee89,Gan93}
have studied another class of higher-dimensional non-Fermi liquids,
where the {\it{current-current
interaction}} mediated by transverse gauge fields
is responsible for the non-Fermi liquid behavior.
Our results are not applicable to these
more complicated interactions.
In contrast
to the density-density interactions discussed here, for 
current-current interactions
it is not allowed to
ignore the curvature of the Fermi surface in a bosonization
approach. Although we know\cite{Kopietz97,Kopietz96}
how to include
curvature effects into the bosonization
calculation of the
single-particle Green´s function,  
it is not straightforward to generalize this method
for the polarization.

Finally, in Sec.\ref{sec:disorder} we have shown that in the perturbative
treatment of disorder in higher-dimensional
Luttinger liquids it is very important to distinguish between
the single-particle scattering time $\tau_1 ( \omega )$ and the
transport scattering time $\tau_{\rm tr} ( \omega )$, 
{\it{even if the
impurity potential is dominated by $s$-wave scattering.}} In particular, 
a possible divergence of $\tau_1 ( \omega )$ 
does not necessarily indicate a similar divergence of the 
conductivity or the
diffusion coefficient.

\section*{Acknowledgements}
We would like to thank Sudip
Chakravarty for discussions and
the Deutsche Forschungsgemeinschaft
for financial support.

%

%
%
\end{document}